\documentclass[aps,prd,twocolumn,aps,floatfix,showpacs,tightenlines,showkeys,superscriptaddress,amsmath,amssymb,nofootinbib]{revtex4-1}
\usepackage{amssymb,amsbsy,epsfig,color,graphicx}
\usepackage{color}
\usepackage{[longtable}
\usepackage{array}
\usepackage{dcolumn}   
\usepackage{cellspace}
\usepackage{mathtools}
\usepackage{amstext}
\usepackage{amssymb}
\usepackage{stmaryrd}
\usepackage{stackrel}
\usepackage{graphicx}
\usepackage{esint}
\usepackage[utf8]{inputenc}
\usepackage{blindtext}
\usepackage{float}
\restylefloat{table}
\usepackage{booktabs}
\usepackage{enumitem} 

\usepackage{etoolbox} 
\usepackage{lipsum} 
\usepackage[capitalize]{cleveref}

\usepackage{multirow}
\usepackage[caption=false]{subfig}

\makeatletter

\appto{\appendix}{%
	\@ifstar{\def\theequation@prefix{A.}}%
	{}%
}
\makeatother

\begin{document}

\title{Quantum spreading of a self-gravitating wave-packet in singularity free gravity }

\author{Luca Buoninfante}
\affiliation{Dipartimento di Fisica "E.R. Caianiello", Universit\`a di Salerno, I-84084 Fisciano (SA), Italy}
\affiliation{INFN - Sezione di Napoli, Gruppo collegato di Salerno, I-84084 Fisciano (SA), Italy}
\affiliation{Van Swinderen Institute, University of Groningen, 9747 AG, Groningen, The Netherlands}

\author{Gaetano Lambiase}
\affiliation{Dipartimento di Fisica "E.R. Caianiello", Universit\`a di Salerno, I-84084 Fisciano (SA), Italy}
\affiliation{INFN - Sezione di Napoli, Gruppo collegato di Salerno, I-84084 Fisciano (SA), Italy}

\author{Anupam Mazumdar}
\affiliation{Van Swinderen Institute, University of Groningen, 9747 AG, Groningen, The Netherlands}
\affiliation{Kapteyn Astronomical Institute, University of Groningen, 9700 AV, Groningen, The Netherlands}

\date{\today}


\begin{abstract}
In this paper we will study for the first time how the wave-packet of a self-gravitating meso-scopic system spreads in theories beyond Einstein's general relativity. In particular, we will consider a ghost-free infinite derivative gravity, which resolves the $1/r$ singularity in the potential - such that the gradient of the potential vanishes within the scale of non-locality. We will show that a quantum wave-packet spreads faster for a ghost-free and singularity-free gravity as compared to the Newtonian case, therefore providing us a unique scenario for testing classical and quantum properties of short-distance gravity in a laboratory in the near future.    
\end{abstract}

\maketitle


\section{Introduction}

On large distances and late times the gravitational interaction is well described by the theory of general relativity (GR) that, indeed, has been very successful since Einstein's initial work, being tested to a very high precision in the infrared (IR)~\cite{-C.-M.}. The most recent success of GR comes from the observation of gravitational waves from merging of binary blackholes which gave a further confirmation of its predictions~\cite{-B.-P.}. Despite these great achievements, our knowledge of the gravitational interaction in the ultraviolet (UV) is still very limited: suffice to say that the inverse-square law of the Newtonian potential has been tested only up to $5.6\times10^{-5}\,\text{m}$ in torsion-balance experiments so far \cite{-D.-J.}. This means that any modification from the Newtonian $1/r$-fall is expected to happen in the large range of values going from the lower bound $0.004\,\text{eV}$ to the Planck scale $M_{p}\sim10^{19}\,\text{GeV}$. This is the place where nature should manifest a different behaviour compared to GR and where either quantum or classical modification from GR should appear in order to solve problems that still remain unsolved as, for example, blackhole and cosmological singularities that make Einstein's theory incomplete in the UV. There have been many theoretical attempts that try to modify GR in the UV regime but none of them have been sufficiently satisfactory so far. In fact, only the experiment will be able to tell us whether the gravitational interaction is really quantum or not, and, in both cases, whether the classical properties are also modified.

Recently, a $\it new$ scenario has been proposed in which by studying the quantum spread of the solitonic wave-packet for a self-gravitating meso-scopic system one can test and constrain modified theories of gravity in the near future \cite{l.buoninfante-solitonic}. In this framework, the so called $\it{infinite}$ $\it{derivative}$ $\it{gravity}$ (IDG) \cite{Biswas:2011ar} was considered as example of alternative theory: it belongs to the class of $\it{non}$-$\it{local}$ ghost-and singularity-free theories of gravity. It was shown that in the non-relativistic and in the weak-field regimes the dynamics of the matter-sector is governed by a Schr\"odinger equation with a non-linear self-interaction term \cite{l.buoninfante-solitonic}:
\begin{equation}
\begin{array}{ll}
{\displaystyle i\frac{\partial}{\partial t}\psi(\vec{x},t)=\left[-\frac{1}{2m}\nabla^{2}\right.}\displaystyle-Gm^{2}\int d^{3}x'\\
\begin{array}{rr}
& \,\,\,\,\,\displaystyle\left.\times\frac{{\rm Erf}\left(\frac{M_{s}}{2}\left|\vec{x}'-\vec{x}\right|\right)}{\left|\vec{x}'-\vec{x}\right|}\left|\psi(\vec{x}',t)\right|^{2}\right]\psi(\vec{x},t),\end{array}
\end{array}\label{eq:1}
\end{equation}
where $G=1/M_{p}^{2}$ and $M_s$ represents the scale of new physics at which non-locality-effects should become relevant, i.e. $0.004{\rm eV} \leq M_s\leq 10^{19}$~GeV. 

Such an integro-differential equation can have two completely different physical interpretations, correspondingly the wave-function $\psi(\vec{x},t)$ can assume two different meanings\footnote{See Refs. \cite{hu,Bahrami} for more details and a review on these two different physical approaches in the case of Newtonian gravity, where the main equation is the Schr\"odinger-Newton equation.}:
\begin{enumerate}
	\item It can appear when gravity is quantized and directly coupled to the stress-energy tensor operator. In this case, it is derived as a Hartree equation in a mean-field approximation; $\psi(\vec{x},t)$ has the meaning of wave-function associated to an $N$-particle state, with large number of particles ($N\rightarrow\infty$), i.e. a condensate \cite{l.buoninfante-solitonic}.
	\item Moreover, Eq. \eqref{eq:1} can be seen as a fundamental equation describing the dynamics of a self-gravitating one-particle system, when considering a semi-classical approach where gravity is coupled to the expectation value of the stress-energy tensor; in this case $\psi(\vec{x},t)$ represents a one-particle wave-function. In such semi-classical framework gravity is treated as a classical interaction, while matter is quantized.
\end{enumerate}
Note that, in case 1. the non-linearity emerges when considering the limit of large number of particles, i.e. in the mean-field regime; while in 2. one has non-linearity even for a one-particle state, bringing to a modified Schr\"odinger equation.

In this respect, a semi-classical approach to IDG would seem more speculative as, not only we would modify GR, but also quantum mechanics. While considering the case of quantized gravity, and studying the dynamics of a self-gravitating condensate, could be particularly more interesting as the main motivations of IDG concern problems emerging when one tries to quantize gravity. However, in this manuscript we wish to make a general treatment by considering both cases of semi-classical and quantized gravitational interaction, and discuss the experimental feasibility of the model in both cases.

It is also worth emphasizing that the non-linear potential term in Eq. \eqref{eq:1} can be split in two parts as follow \cite{l.buoninfante-solitonic}:
\begin{equation}
\begin{array}{rl}
\!\!\!V[\psi](\vec{x})\simeq &\displaystyle -\frac{G m^2 M_s}{\sqrt{\pi}}\!\!\!\int\limits_{|\vec{x}'|<2/M_s}d^3 x' \left|\psi(\vec{x}',t)\right|^2\\
& -Gm^2 \displaystyle \!\!\!\int\limits_{|\vec{x}'|\geq 2/M_s} d^3 x' \frac{\left|\psi(\vec{x}',t)\right|^2}{|\vec{x}'-\vec{x}|}.
\end{array} \label{decomp}
\end{equation}
From the last decomposition one can notice that the first term contains all information about the non-local nature of the gravitational interaction, while the second one has the same form of the usual Newtonian self-potential that also appears in the well-known Schr\"odinger-Newton equation, see \cite{Diosi:1988uy,Penrose,Bahrami}.

In Ref. \cite{l.buoninfante-solitonic} it was shown that Eq. \eqref{eq:1} admits stationary solitonic-like solutions for the ground-state and it was found that in the case of IDG the energy $E$ and the spread $\sigma$ of the solitonic wave-packet turn out to be $\it larger$ compared to the respective ones in Newtonian gravity, i.e. $E_{{\scriptscriptstyle IDG}}\geq E_{{\scriptscriptstyle N}}$ and $\sigma_{{\scriptscriptstyle IDG}}> \sigma_{{\scriptscriptstyle N}}$; these are effects induced by the non-local nature of the gravitational interaction \cite{l.buoninfante-solitonic}. 

The expression of the ground-state energy is given by
\begin{equation}
E_{{\scriptscriptstyle IDG}}=\frac{3}{4}\frac{1}{m\sigma^{2}}-\sqrt{\frac{2}{\pi}}Gm^{2}\frac{M_{s}}{\sqrt{2+M_{s}^{2}\sigma^{2}}},\label{eq:2}
\end{equation}
that in the limit  $M_{s}\sigma>2$ recovers the energy of Newton's theory $E_{{\scriptscriptstyle N}}=\frac{3}{4}\frac{1}{m\sigma^{2}}-\sqrt{\frac{2}{\pi}}\frac{Gm^{2}}{\sigma}$ \cite{Diosi:1988uy}. The above Eq. \eqref{eq:2} shows the action of two kinds of $\it{forces}$ that are completely different in nature: a quantum-mechanical kinetic contribution that tends to spread the wave-packet and the gravitational potential which takes into account the attractiveness of gravity coming from the non-linear term of Eq. \eqref{eq:1}. In a stationary scenario the two contributions balance each other and the soliton-like solution above can be found.

In this paper, unlike Ref. \cite{l.buoninfante-solitonic}, we are more interested in studying $\it non$-$\it stationary$ solutions of Eq. \eqref{eq:1}. In particular we want to understand how the $\it{spreading}$ of the wave-packet is affected by the presence of a $\it non$-$\it local$ gravitational self-interaction. The analysis that we will present will apply to both cases of semi-classical and quantized gravity as the main equation is mathematically the same.

As pointed out in Refs. \cite{harrison,carlip,giulini}\footnote{In Refs. \cite{harrison,carlip,giulini} the authors mainly focused on the semi-classical approach, where gravity is treated classically. However, the following treatment will also apply to the case of quantized gravity as one has to consider the same integro-differential equation \eqref{eq:1}. Let us keep in mind that in the semi-classical approach the quantum wave-packet represents a one-particle wave-function; while, when gravity is quantized, it is associated to the dynamics of a many-particle system, i.e. a condensate.}, where numerical studies of the Schr\"odinger-Newton equation were made, there should exist a threshold mass $\mu$, such that the collapse of the wave-function induced by gravity will take place for any $m>\mu$. In Ref. \cite{harrison} it was noticed that the collapsing behavior appears {\it only} if the initial state of the quantum system has negative energy, such that the attractive contribution of self-gravity dominates. From this last observation, we understand that a possible way to find an analytical estimation for the threshold mass $\mu$ is to equate kinetic and gravitational contributions in Eq. \eqref{eq:2}; thus we obtain
\begin{equation}
\mu_{\scriptscriptstyle IDG}=\left(\frac{3}{4}\sqrt{\frac{\pi}{2}}\frac{\sqrt{2+M_{s}^{2}\sigma^{2}}}{G\sigma^{2}M_s}\right)^{\frac{1}{3}}.\label{eq:3} 
\end{equation}
Note that in the limit when $M_s \sigma > 2$, Eq. \eqref{eq:3} gives the threshold mass similar to the case of Newtonian theory:
\begin{equation}
\mu_{\scriptscriptstyle N}=\left(\frac{3}{4}\sqrt{\frac{\pi}{2}}\frac{1}{G\sigma}\right)^{\frac{1}{3}}.\label{eq:4} 
\end{equation}
Eqs. \eqref{eq:3} and \eqref{eq:4} clearly show that non-locality implies a larger value of the threshold mass, i.e. $\mu_{\scriptscriptstyle IDG}>\mu_{\scriptscriptstyle N}$ for any values of $M_s$ and $\sigma$. If we choose the current lower bound on the scale of non-locality, $M_s=0.004$ eV \cite{Edholm:2016hbt}, and $\sigma=500$ nm,%
\footnote{It is worthwhile to note that the special choice we have made for the value of the spread, $\sigma=500$ nm, corresponds to the actual slit separation $d$ in a Talbot-Lau interferometry setup~\cite{stibor}. The slit separation is related to both length $L$ of the device and de Broglie wave-length $\lambda =h/mv$, where $h$ is the Planck's constant, $m$ the mass of the particle and $v$ is its velocity, through the relation $L=d^{2}/\lambda$. The interference with larger masses requires smaller wavelengths, which in turn means shorter slit-separations. Moreover, the most massive quantum systems which have been seen showing interference are organic molecules with a mass of the order of $10^{-22}$-$10^{-21}$ kg \cite{arndt2005}. \label{foot:1} %
}
the values of the threshold masses are $\mu_{\scriptscriptstyle IDG}\simeq 3.5\times 10^{-17}$ kg and $\mu_{\scriptscriptstyle N}\simeq6.7\times 10^{-18}$ kg.

From an analytical estimation one expects that for masses, $m>\mu_{\scriptscriptstyle IDG}$, the self-gravitating quantum wave-packet collapses, while for masses, $m<\mu_{\scriptscriptstyle IDG}$, one would expect no collapse of the wave-packet, but only a slow-down of the spreading compared to that of the free-particle case. 

We now wish to study the quantum spreading of a self-gravitating wave-packet and understand how it is affected by singularity-free gravity, without taking into account the collapsing phase. It means we will work in a regime in which we can assume that the non-linear contribution in Eq. \eqref{eq:1} is smaller compared to the kinetic term:
\begin{equation}
\sqrt{\frac{2}{\pi}}\frac{Gm^{2}M_{s}}{\sqrt{2+M_{s}^{2}\sigma^{2}}}<\frac{3}{4}\frac{1}{m\sigma^{2}}.\label{eq:5} 
\end{equation}
In this regime non-linearity-effects are sufficiently small and it allows us to find non-stationary solutions by applying the Fourier analysis to Eq. \eqref{eq:1}. From Eq. \eqref{eq:5}, we can also define the dimensionless parameter 
\begin{equation}
\xi:=\frac{Gm^{3}\sigma^{2}M_s}{\sqrt{2+M_{s}^2\sigma^{2}}}, \label{param-xi}
\end{equation}
that quantifies the $\it degree$ of non-linearity due to self-gravity. It depends on the initial data through the mass $m$ and $\sigma$, that can represent the initial spread of the self-gravitating quantum wave-packet. When $\xi <1$, we can assume that the non-linear effects are sufficiently small. In the Newtonian limit, $M_s \sigma_{0} > 2$, we will obtain $\xi \sim Gm^{3}\sigma$.

A comparison between modified theories of gravity, IDG in our case, and Newton's gravity, will provide us a new and unique framework to test short-distance gravity beyond Einstein's GR.
This paper is organized as follows: first of all we will briefly introduce ghost-free and singularity-free IDG; then we will study the spreading solutions of Eq. \eqref{eq:1} with the aim of comparing free, Newton and IDG cases; finally there will be a summary and a discussion on current and near future experimental scenarios in both cases of semi-classical and quantized gravity.    

\section{Infinite derivative ghost-free and singularity-free gravity}  

There have been many attempts to modify GR by introducing higher order derivative contribution in the action, especially a conformal gravity containing quadratic terms in the curvature like $\mathcal{R}^{2},$ $\mathcal{R}_{\mu\nu}\mathcal{R}^{\mu\nu},$ $\mathcal{R}_{\mu\nu\rho\sigma}\mathcal{R}^{\mu\nu\rho\sigma}.$ Such quadratic theory of gravity turns out to be conformal as well as renormalizable, but it suffers from the presence of a massive spin-$2$ ghost field that makes the theory classically unstable and non-unitary at the quantum level \cite{-K.-S.}. 

Recently, it has been noticed that by considering an $\it infinite$ number of derivatives in the quadratic curvature gravitational action one can prevent the presence of ghost~\cite{Biswas:2011ar,Biswas:2005qr}. At the same time, such a ghost-free action also improves the behaviour of the gravitational interaction in the UV regime showing a non-singular potential and a vanishing gravitational force: $\Phi \rightarrow \text{const}$ and $F_g \rightarrow 0$ as $ r\rightarrow 0$, where $F_g$ represents the mutual force between two particles separated by the distance $r$ \cite{Biswas:2011ar}. 

The most general torsion-free, parity-invariant and quadratic covariant action that contains an infinite number of derivatives has been constructed around constant curvature backgrounds, and reads \cite{Biswas:2011ar,Biswas:2016etb}:
\begin{equation}
\begin{array}{rl}
S= & \displaystyle \frac{1}{16\pi G}\int d^{4}x\sqrt{-g}\left[{\displaystyle \mathcal{R}}+\alpha\left(\mathcal{R}\mathcal{F}_{1}(\boxempty_{s})\mathcal{R}\right.\right.\\
& \begin{array}{ll}
\left.\left.+\mathcal{R}_{\mu\nu}\mathcal{F}_{2}(\boxempty_{s})\mathcal{R}^{\mu\nu}+\mathcal{R}_{\mu\nu\rho\sigma}\mathcal{F}_{3}(\boxempty_{s})\mathcal{R}^{\mu\nu\rho\sigma}\right)\right],\end{array}
\end{array}\label{eq:6}
\end{equation}
where $\alpha$ is a dimensionful coupling, $\boxempty_{s}\equiv\boxempty/M_{s}^{2}$ and $\boxempty\equiv g^{\mu\nu}\nabla_{\mu}\nabla_{\nu}$, where $\mu,~\nu=0,1,2,3$, and the mostly positive metric signature, $(-,+,+,+)$, is chosen. The information about the presence of infinite derivatives is contained in the three gravitational form factors $\mathcal{F}_{i}(\boxempty_{s})$ which have to be analytic functions of $\boxempty$, $\mathcal{F}_{i}(\boxempty_{s})=\stackrel[n=0]{\infty}{\sum}f_{i,n}\left(\boxempty_{s}\right)^{n}$, thus we can smoothly recover GR when we take the limit $\boxempty \rightarrow 0$. These form factors can be further constrained by requiring general covariance, that no additional dynamical degrees of freedom propagate other than the massless graviton and the ghost-free condition, that around Minkowski background is given by $2{\cal F}_1(\boxempty_s)+ {\cal F}_2(\boxempty_s) + 2{\cal F}_3(\boxempty_s)=0$~\cite{Biswas:2011ar}. Note that around a constant curvature spacetime we can set ${\cal F}_3=0$ \cite{Biswas:2011ar}, without loss of generality. 

As we have already mentioned above, the parameter $M_s$ represents the scale of non-locality where gravity, described by this class of ghost- and singularity-free theories, shows a non-local nature \cite{-Yu.-V.,Tomboulis,Tseytlin,Siegel,Modesto,Biswas:2005qr,Talaganis:2014ida,frolov}. The current constraints on $M_s$ comes from torsion-balance experiments which have seen no departure from the Newtonian $1/r$-fall up to a  distance of $5.6\times 10^{-5}$ meters, that implies $M_s \geq 0.004$ eV \cite{Edholm:2016hbt,Perivolaropoulos:2016ucs}.

Furthermore, it is worth noting that IDG-theory can also resolve cosmological singularity, see~\cite{Biswas:2005qr,Biswas:2011ar,Koshelev:2012qn}, and the non-local nature of gravity can possibly even play a crucial role in the resolution of blackhole singularity as pointed out in Ref.~\cite{Koshelev:2017bxd}; while at the quantum level it is believed that the action in Eq. \eqref{eq:6} describes a gravitational theory that is UV-finite beyond 1-loop~\cite{-Yu.-V.,Tomboulis,Modesto,Talaganis:2014ida}.

The ghost-free condition of IDG demands a special choice for the gravitational form factors, see \cite{Biswas:2005qr,Biswas:2011ar}:\footnote{The exponential choice $e^{-\boxempty/M_s^{2}}$ is made in order to have a UV-suppression in the propagator in momentum space. Indeed, the dressed physical propagator turns out to be suppressed either for time-like and space-like momentum exchange \cite{okada}. Note that if we had assumed to work with the mostly negative signature, we would have had to choose $e^{\boxempty/M_s^{2}}$. In both cases one obtains a well-defined gravitational potential that recovers the correct Newtonian limit in the IR. Both choices are also compatible with the change of sign of the kinetic term in the graviton Lagrangian depending on the signature convention, $h_{\mu \nu}\boxempty h^{\mu \nu}$ and $-h_{\mu \nu}\boxempty h^{\mu \nu}$, respectively, where $h_{\mu \nu}$ is the graviton field defined as metric perturbation around flat spacetime, $g_{\mu \nu}=\eta_{\mu \nu}+h_{\mu \nu}$. Moreover, integrations in momentum space with such exponentials can be performed by following various prescriptions as, for example, Wick rotation to Euclidean space, or for alternative prescriptions see also \cite{Talaganis:2014ida,okada} and \cite{sen}.}
\begin{equation}
\alpha {\cal F}_{1}(\boxempty_{s})=-\frac{\alpha}{2}{\cal F}_{2}(\boxempty_{s})=\frac{a(\boxempty_{s})-1}{\boxempty},\,\,\,\,a(\boxempty_{s})=e^{-\boxempty/M_s^{2}}.\label{eq:7}
\end{equation}
Generally, $a(\boxempty_{s})$ should be {\it exponential of an entire function}~\cite{Tomboulis,Biswas:2005qr,Biswas:2011ar,Modesto}, in order to avoid additional dynamical degrees of freedom other than the massless spin-$2$ graviton, therefore {\it no} propagating {\it ghost}-like states. In fact, any generalised form of exponential of an entire function yields a similar behavior in the UV and IR regimes, namely a similar non-singular modified gravitational potential that for large distances recovers the Newtonian $1/r$-fall \cite{Edholm:2016hbt,frolov-zelnikov}.

By linearizing the action in Eq. \eqref{eq:6} and going to momentum space one can easily show that the choice in Eq. \eqref{eq:7} does not introduce any extra degrees of freedom in the gravity sector. Indeed, as shown in Ref. \cite{Biswas:2011ar,Biswas:2013kla,Buoninfante}, the gauge-independent part of the propagator corresponding to the linearized action around Minkowski spacetime is given by
\begin{equation}
\Pi(-k^2)=\frac{1}{a(-k^2)}\left(\frac{\mathcal{P}^{2}}{k^{2}}-\frac{\mathcal{P}^{0}}{2k^{2}}\right),\label{eq:8}
\end{equation}
where $\mathcal{P}^{2}$ and $\mathcal{P}^{0}$ are the well known spin projector operators that project any symmetric two-rank tensor along the spin-$2$ and spin-$0$ components, respectively; $\Pi_{{\scriptscriptstyle GR}}=\mathcal{P}^{2}/k^{2}-\mathcal{P}_{s}^{0}/2k^{2}$ is the GR propagator. For the special choice $a(-k^{2})=e^{k^{2}/M_{s}^{2}}$, there are no additional poles in the complex plane and thus only the massless graviton propagates.

We are interested in the non-relativistic, weak-field and static spacetime approximations, such that we can compute the gravitational potential from which, in turn, one can write down the Hamiltonian interaction coupling gravity and matter sectors in both cases of semi-classical and quantized gravitational interaction, as it has been done in Ref. \cite{l.buoninfante-solitonic}. 

As shown in Ref. \cite{l.buoninfante-solitonic}, in the semi-classical approach gravity is coupled to the expectation value of the quantum energy-stress tensor, and the field equation for the potential, with the choice Eq. \eqref{eq:7}, reads\footnote{Since terms with derivatives of order higher than four are usually neglected at small curvature, one could be brought to think that the Taylor expansion of the exponential $e^{-\nabla^{2}/M_{s}^{2}}$ can be truncated. However, it is not the case here: in fact, in Eq. \eqref{eq:6} we are considered the most general quadratic-curvature action, and the infinite-order in derivatives comes from the form factors ${\cal F}_i(\Box)$ and not from higher order curvature-invariants. Moreover, even if we wanted to truncate the series, we would suffer from the ghost problem again.}:
\begin{equation}
\begin{array}{rl}
e^{-\nabla^{2}/M_{s}^{2}}\nabla^{2}\Phi= & 4\pi G \left\langle \psi \left| \hat{\tau}_{00}\right| \psi\right\rangle\\
= &  4\pi Gm \left| \psi(\vec{x},t)\right|^{2},\label{eq:9}
\end{array}
\end{equation}
whose solution is given by
\begin{equation}
\Phi[\psi](\vec{x})=-Gm\int d^{3}x'\frac{{\rm Erf}\left(\frac{M_{s}}{2}\left|\vec{x}'-\vec{x}\right|\right)}{\left|\vec{x}'-\vec{x}\right|}\left|\psi(\vec{x}',t)\right|^{2},\label{eq:10}
\end{equation}
i.e. one has a classical gravitational potential generated by the probability density $\left|\psi(\vec{x},t)\right|^{2}$ that plays the role of a semi-classical source. In a Hamiltonian formulation the potential in 
Eq. \eqref{eq:10} contributes to the self-interaction of matter as described by the modified Schr\"odinger equation in Eq. \eqref{eq:1}, see \cite{l.buoninfante-solitonic}.

In the case of quantized gravity instead, the graviton field is directly coupled to the quantum energy-stress tensor so that the analog of Eq. \eqref{eq:9} reads
\begin{equation}
e^{-\nabla^{2}/M_{s}^{2}}\nabla^{2}{\hat \Phi}=  4\pi G \hat{\tau}_{00}, \label{eq:9.2}
\end{equation}
whose solution is given by
\begin{equation}
{\hat \Phi}(\vec{x})=-Gm\int d^{3}x'\frac{{\rm Erf}\left(\frac{M_{s}}{2}\left|\vec{x}'-\vec{x}\right|\right)}{\left|\vec{x}'-\vec{x}\right|}{\hat \psi}^{\dagger}(\vec{x}'){\hat \psi}(\vec{x}').\label{eq:10.2}
\end{equation}
Note that we have used ${\hat \tau}_{00}\equiv{\hat \rho}=m\psi^{\dagger}\psi.$ By calculating also in this case the Hamiltonian interaction, it becomes clear that the quantum gravitational potential in Eq. \eqref{eq:10.2} does not introduce any non-linearity in a $N$-particle Schr\"odinger equation, but the non-linear integro-differential equation \eqref{eq:1} would emerge when considering mean-field regime for the many-body system ($N\rightarrow\infty$) \cite{l.buoninfante-solitonic}.

In the following section we will study the spreading solutions of Eq. \eqref{eq:1} and, by comparing to the case of Newtonian gravity, we will be able to see which is the effect of non-locality on a self-gravitating wave-packet in IDG-theory. The analysis will hold for both cases of semi-classical and quantized gravity as the main dynamical equation is mathematically the same, i.e. Eq. \eqref{eq:1}.

\section{Spreading solutions for a self-gravitating wave-packet}
We now wish to study non-stationary solutions of the non-linear integro-differential equation in Eq. \eqref{eq:1} by working in a regime where non-linearity-effects can be considered sufficiently small such that there will be no gravity-induced collapse. Such a regime is the one described by the inequality in Eq. \eqref{eq:5} which gives the range of masses (see Eqs. \eqref{eq:3} and \eqref{eq:4}) for which the self-gravitating wave-packet will not collapse, as the attractive contribution due to gravity is not dominating, but it can allow $\it only$ the spread of the wave-packet. In this scenario, we are allowed to study Eq. \eqref{eq:1} in the Fourier space.

Let us suppose we start with an initial Gaussian wave-packet:
\begin{equation}
\psi(\vec{x},0)={\displaystyle \frac{1}{\pi^{3/4}\sigma_{0}^{3/2}}e^{-\left|\vec{x}\right|^{2}/2\sigma_{0}^{2}}},\,\,\,\,\int d^{3}x\psi(\vec{x},0)^{2}=1,\label{eq:11}
\end{equation}
where $\sigma_0$ is the initial spread. A formal expression of the wave-packet at a generic time $t>0$ can be found in terms of its Fourier transform:
\begin{equation}
\begin{array}{rl}
\psi(\vec{x},t)= & \displaystyle \int \frac{d^{3}k d\omega}{(2 \pi)^{4}} \phi (\vec{k},\omega) e^{i(\vec{k} \cdot \vec{x}-\omega t)}\\
 = & \displaystyle \int \frac{d^{3}k}{(2 \pi)^{3}} \phi(\vec{k}) e^{i(\vec{k} \cdot \vec{x}-\omega (\vec{k}) t)},
\end{array}\label{eq:12}
\end{equation}
where we have used $\phi (\vec{k},\omega)=2\pi \phi (\vec{k}) \delta (\omega -\omega (\vec{k}))$, and $\phi (\vec{k})$ can be obtained by calculating the anti-Fourier transform at the initial time $t=0$:
\begin{equation}
\begin{array}{rl}
\phi (k)= & \displaystyle \int d^{3}x \psi(\vec{x},0) e^{-i\vec{k} \cdot \vec{x}}\\
= & 2 \sqrt{2} \pi^{3/4} \sigma_{0}^{3/2} e^{-\frac{1}{2}k^{2}\sigma_{0}^{2}}, \label{eq:13}
\end{array}
\end{equation}
where $k \equiv |\vec{k}|$. By using Eq. \eqref{eq:12}, and the expression of the IDG potential in the momentum space, 
\begin{equation}
\frac{\mathrm{Erf}\left(\frac{M_{s}}{2}\left|\vec{x}'-\vec{x}\right|\right)}{\left|\vec{x}'-\vec{x}\right|}= \int \frac{d^{3}k}{(2\pi)^{3}}\frac{4\pi e^{-k^{2}/M_{s}^{2}}}{k^{2}}e^{i\vec{k}\cdot(\vec{x}'-\vec{x})},\label{eq:14}
\end{equation}
and by acting with $\int d^{3}xdt e^{-i(\vec{k}\cdot \vec{x}-\omega t)}$ on both sides of the modified Schr\"odinger equation in Eq. \eqref{eq:1}, we obtain the dispersion-frequency $\omega_{{\scriptscriptstyle IDG}}$ as a function of $k$:
\begin{equation}
\omega_{{\scriptscriptstyle IDG}} (k)=	\frac{k^2}{2m} - 32Gm^{2}\pi^{5/2}\sigma_{0}^{3}{\cal D}_{{\scriptscriptstyle IDG}}(k),\label{eq:15}
\end{equation}
where 
\begin{equation}
\begin{array}{ll}
\mathcal{D}_{\scriptscriptstyle IDG}(k)= & \\
\begin{array}{rr} \displaystyle \int\frac{d^{3}k'd^{3}k''}{(2\pi)^{3}(2\pi)^{3}}\frac{e^{-|\vec{k}''-\vec{k}'|^{2}(\sigma_{0}^{2}+1/M_{s}^{2})}e^{[(\vec{k}''-\vec{k}')\cdot \vec{k}-\vec{k}''\cdot \vec{k}'] \sigma_{0}^{2}}}{|\vec{k}''-\vec{k}'|^2}. &  \label{eq:16}
\end{array}\end{array}
\end{equation}
In order to solve the integral in Eq. \eqref{eq:16} we can make the following change of integration variables: $\vec{X}:=\vec{k}''-\vec{k}',$ $\vec{Y}:=\vec{k}''+\vec{k}'$, thus $\vec{k}''\cdot \vec{k}'=(Y^2-X^2)/4$ and the integral turns out to be decoupled in two other integrals that can be easily calculated by using polar coordinates:%
\footnote{The special function $\mathrm{Erfi}(x)$ is the imaginary error-function and is defined as $\mathrm{Erfi}(x):=\frac{2}{\sqrt{\pi}} \int_{0}^{x} dt e^{t^2}\equiv -i\mathrm{Erf}(ix).$}
\begin{figure}[t]
	\includegraphics[scale=0.445]{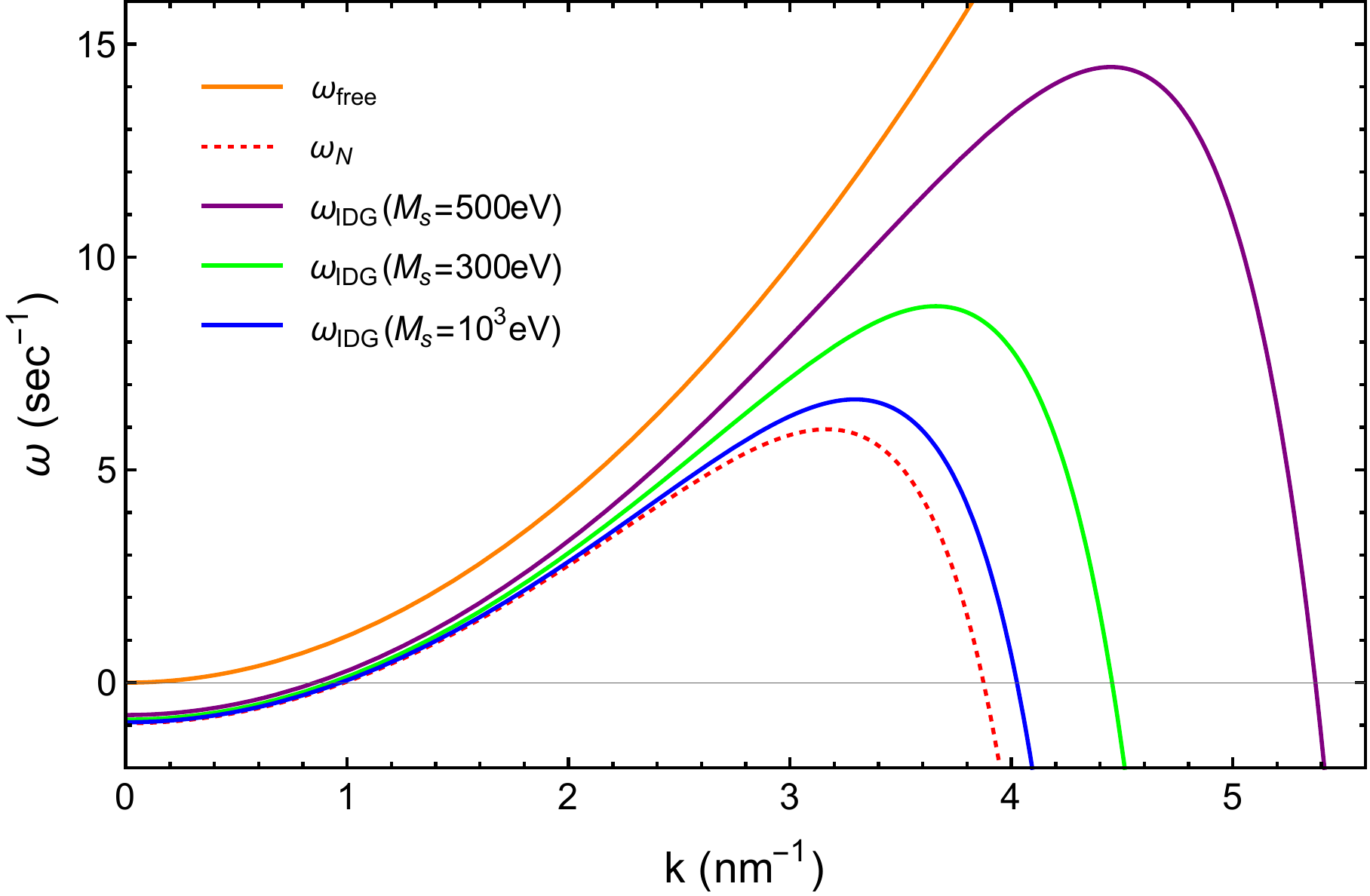}
	\caption{In the above plot we have drawn the dispersion relations (in sec$^{-1}$) for the free particle case   
		$\omega_{\scriptscriptstyle{\text{free}}}$ (orange line), for the Newtonian case
		$\omega_{{\scriptscriptstyle N}}$ (dashed red line), and for the IDG case $\omega_{{\scriptscriptstyle IDG}}$, with respect to the wave-vector $k$ (in nm$^{-1}$). For IDG-theory, we have also considered different values of $M_s= 300\,\text{eV}$ (purple line), 
		$500\,\text{eV}$ (green line) and $10^{3}\,\text{eV}$ (blue line). We have chosen the values $m=4.8\times 10^{-17}$ kg and $\sigma_{0}=1$ nm for the mass and the initial spread, respectively.}\label{fig1}
\end{figure}
\begin{equation}
\begin{array}{rl}
\!\!\!\mathcal{D}_{\scriptscriptstyle IDG}(k)= & \displaystyle \!\!\!\frac{1}{8}\!\int\!\!\!\frac{d^{3}X}{(2\pi)^{3}}\frac{e^{-(\frac{3}{4}\sigma_{0}^{2}+\frac{1}{M_{s}^{2}})|\vec{X}|^2}e^{\vec{X}\cdot \vec{k} \sigma_{0}^{2}}}{|\vec{X}|^2}\!\! \!\int\!\!\!\frac{d^{3}Y}{(2\pi)^{3}}e^{-\frac{|\vec{Y}|^2 \sigma_{0}^{2}}{4}}\\
= & \!\!\displaystyle \frac{1}{32\pi^{5/2} \sigma_{0}^{5}} \frac{1}{k}\mathrm{Erfi}\left(\frac{k M_s \sigma_{0}^{2}}{\sqrt{4+3M_{s}^{2}\sigma_{0}^{2}}} \right), \label{eq:17}
\end{array}
\end{equation}
From Eq. \eqref{eq:17} we obtain an expression for the dispersion relation in Eq. \eqref{eq:15} in the case of IDG self-interaction:
\begin{equation}
\omega_{{\scriptscriptstyle IDG}} (k)=	\frac{k^2}{2m} - \frac{Gm^{2}}{\sigma_{0}^{2}}\frac{1}{k}\mathrm{Erfi}\left(\frac{k M_s \sigma_{0}^{2}}{\sqrt{4+3M_{s}^{2}\sigma_{0}^{2}}} \right),\label{eq:18}
\end{equation}
note that in the regime $M_s \sigma_{0}>2$, gives the corresponding Newtonian limit:
\begin{equation}
\omega_{{\scriptscriptstyle N}} (k)=	\frac{k^2}{2m} - \frac{Gm^{2}}{ \sigma_{0}^{2}}\frac{1}{k}\mathrm{Erfi}\left(\frac{k\sigma_{0}}{\sqrt{3}} \right).\label{eq:19}
\end{equation}
In Fig. \ref{fig1} it is shown the behavior of the dispersion relation $\omega (k)$ in the free, Newtonian and IDG cases, and one can immediately notice that as the parameter $M_s$ increases the frequency $\omega_{{\scriptscriptstyle IDG}}$ tends to $\omega_{{\scriptscriptstyle N}}$.

The dispersion relation in Eq. \eqref{eq:18} is crucial in order to determine the time evolution of the wave-packet, indeed from Eq. \eqref{eq:12} the solution $\psi_{{\scriptscriptstyle IDG}} (\vec{x},t)$ is expressed in terms of the frequency $\omega_{{\scriptscriptstyle IDG}}$:
\begin{equation}
\psi_{{\scriptscriptstyle IDG}}(\vec{x},t)= 2\sqrt{2}\pi^{3/4} \sigma_{0}^{3/2} \!\!\!\int \!\!\frac{d^{3}k}{(2 \pi)^{3}} e^{-\frac{1}{2}k^{2}\sigma_{0}^{2}}  e^{i(\vec{k} \cdot \vec{x}-\omega_{{\scriptscriptstyle IDG}}(k) t)}. \label{eq:20}
\end{equation}
The integral in Eq. \eqref{eq:20} cannot be solved analytically, but we will be able to find numerical solutions. 

First of all, note that in the free-particle case there exists a well know analytical solution that describes a quantum-mechanical spreading of the wave-packet, and the corresponding probability density reads:
\begin{equation}
\left|\psi_{\scriptscriptstyle{\text{free}}}(\vec{x},t)\right|^2= \frac{1}{\pi^{3/2}\sigma_{0}^{3}}\frac{e^{-\frac{|\vec{x}|^{2}}{\sigma_{0}^{2}\left(1+\left(\frac{t}{m\sigma_{0}^{2}}\right)^{2}\right)}}}{\left(1+\left(\frac{t}{m\sigma_{0}^{2}}\right)^{2}\right)^{3/2}}, \label{eq:21}
\end{equation}
from which one can see that there exist a time-scale for the spreading, i.e. a time after which the particle turns out to be de-localized, and it is given by:
\begin{equation}
\tau_{\scriptscriptstyle{\text{free}}}=m \sigma_{0}^{2}. \label{eq:22}
\end{equation}

Note that the time-scale in Eq. \eqref{eq:22} can be also obtained by imposing the equality $\tau_{\scriptscriptstyle{\text{free}}}=1/2\omega_{\scriptscriptstyle{\text{free}}}(1/\sigma_{0})$. We argue that in the same way we can also obtain an analytical estimation for the spreading time-scale of a self-gravitating system, thus by using Eq. \eqref{eq:18} and imposing $\tau_{\scriptscriptstyle IDG}=1/2\omega_{\scriptscriptstyle IDG}(1/\sigma_{0})$ we obtain\footnote{Although an exact analytical derivation is lacking for such a time-scale, our argument turns out to be consistent with the numerical analyses, some of which are presented in the end of this section.}
\begin{equation}
\tau_{\scriptscriptstyle{IDG}}\sim \frac{m\sigma_{0}^{2}}{1-2Gm^{3}\sigma_{0}\mathrm{Erfi}\left(\frac{M_s \sigma_{0}}{\sqrt{4+3M_{s}^{2}\sigma_{0}^{2}}} \right)}, \label{eq:23}
\end{equation}
that in the case of Newtonian self-interaction reduces to
\begin{equation}
\tau_{\scriptscriptstyle{N}}\sim \frac{m\sigma_{0}^{2}}{1-2Gm^{3}\sigma_{0}\mathrm{Erfi}\left(\frac{1}{\sqrt{3}} \right)}. \label{eq:time-scale Newton}
\end{equation}
In the opposite regime $M_s \sigma_{0}<2$, when non-locality-effects become dominant, the time-scale in Eq. \eqref{eq:23} assume the following form:
\begin{equation}
\tau_{\scriptscriptstyle{IDG}}^{\scriptscriptstyle{(M_s \sigma_{\scriptscriptstyle 0}<2)}}\sim \frac{m\sigma_{0}^{2}}{1-\frac{2Gm^{3}M_s \sigma_{0}^{2}}{\sqrt{\pi}}},
\end{equation}
that is the same time-scale that we would have if there was a constant gravitational potential.
\begin{figure}[b]
	\centering
	\subfloat[Subfigure 1 list of figures text][$t=0.4$ sec]{
		\includegraphics[scale=0.38]{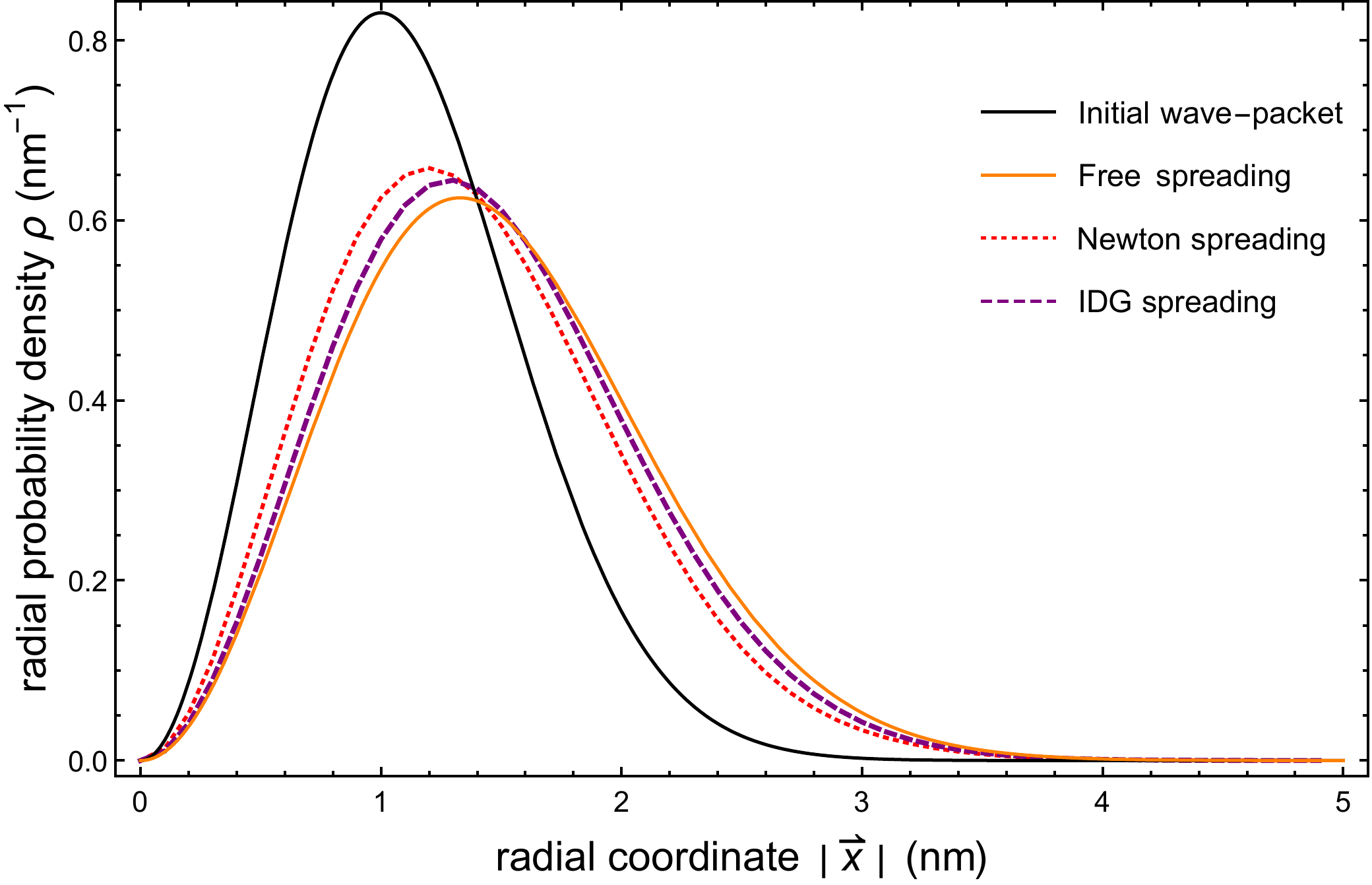}
		\label{fig:subfig1}}
	\qquad
	\subfloat[Subfigure 2 list of figures text][$t=0.8$ sec]{
		\includegraphics[scale=0.38]{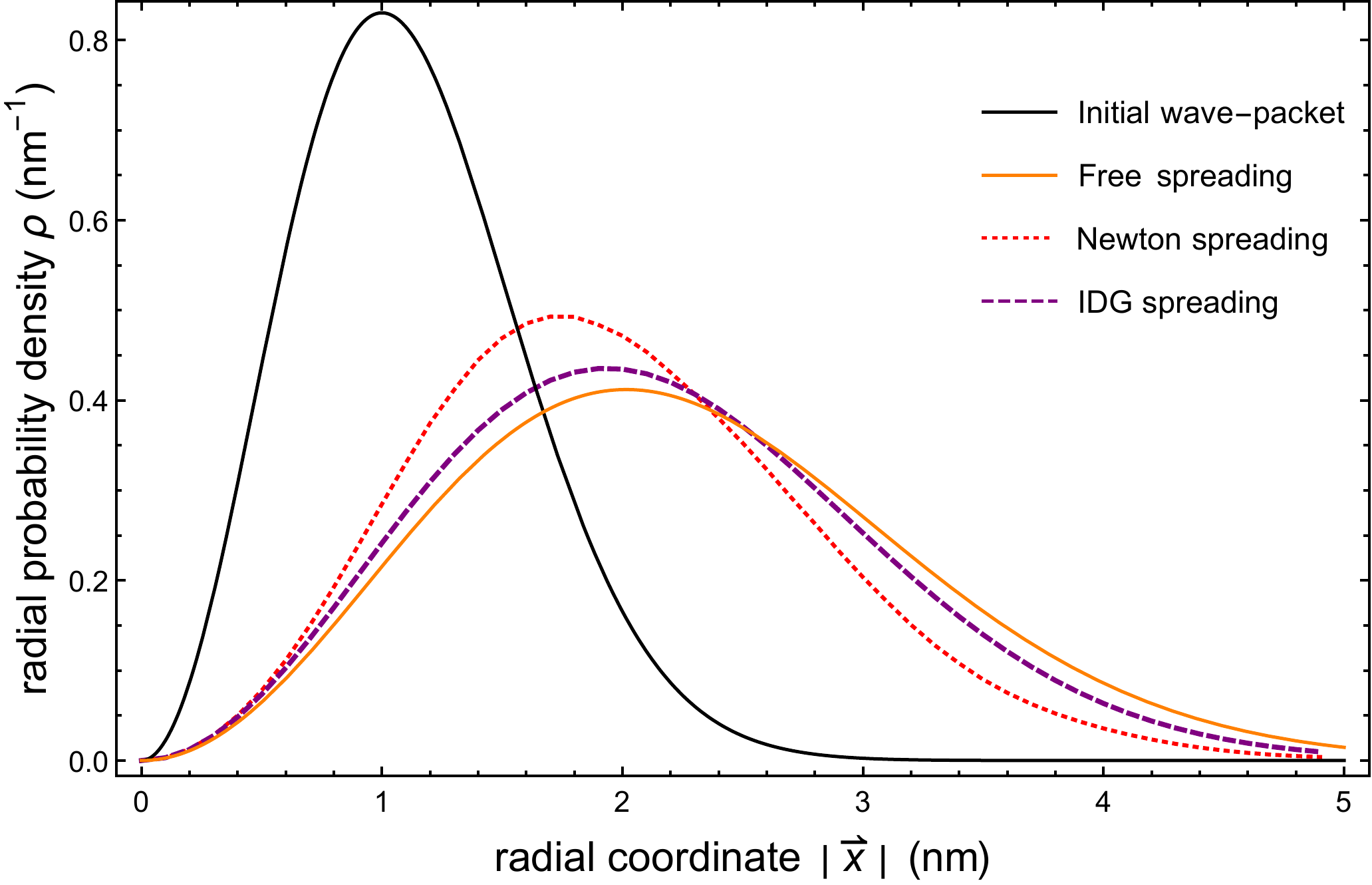}
		\label{fig:subfig2}}
	\protect\caption{In the above plots we have shown the radial probability density $\rho(\vec{x},t)$ (in nm$^{-1}$) as a function of the radial coordinate $|\vec{x}|$ (in nm) at two different fixed times (a) $t=0.4$ sec and (b) $t=0.8$ sec, in the cases of free-particle (orange line), Newton's gravity (dotted red line) and IDG (dashed purple line). The initial probability density $\rho(\vec{x},0)$ is represented by the black line. We have chosen the values $m=4.8\times 10^{-17}$ kg, $\sigma_{0}=1$ nm and $M_s=300$ eV for the mass, the initial spread and the scale of non-locality, respectively.}\label{fig2}
\end{figure}

\begin{table*}[t!]
	\caption{Values of the time-scales $\tau_{\scriptscriptstyle{\text{free}}}$, $\tau_{{\scriptscriptstyle N}}$ and $\tau_{{\scriptscriptstyle IDG}}$ of the spreading wave-packet (in sec) for fixed values of the initial spread $\sigma_{0}$ (in nm), of the mass $m$ (in kg), chosen of the order of the threshold $\chi_{\scriptscriptstyle N}$, and for sample values of $M_s=1\,{\rm eV}, 300\,{\rm eV}, 10^{3}\,{\rm eV}$.}
	\centering
	\begin{tabular}{p{0.10\linewidth}p{0.15\linewidth}p{0.13\linewidth}p{0.13\linewidth}p{0.13\linewidth}p{0.13\linewidth}p{0.13\linewidth}}
		\toprule[1pt]\midrule[0.3pt]
		$\sigma_{0}$  & $m\sim \chi_{\scriptscriptstyle N}$ & $\tau_{\scriptscriptstyle{\text{free}}}$ &  $\tau_{\scriptscriptstyle{N}}$ & $\tau_{{\scriptscriptstyle IDG}}(1$ eV) & $\tau_{{\scriptscriptstyle IDG}}(300$ eV) & $\tau_{\scriptscriptstyle{IDG}}(10^{3}$ eV)\\
		\midrule[0.3pt]
		$500$ nm & $6.0\times10^{-18}$ kg & $14285.7$ sec & $325995$ sec & $97105.2$ sec  &  $325985$ sec & $325994$ sec \\
		$100$ nm & $1.0\times10^{-17}$ kg & $952.381$ sec & $8306.86$ sec & $1407.93$ sec  &  $8304.58$ sec & $8306.65$ sec  \\
		$10$ nm & $2.2\times10^{-17}$ kg & $20.9524$ sec & $365.794$ sec & $21.756$ sec  &  $345.655$ sec & $363.876$ sec  \\
		$1$ nm & $4.8\times10^{-17}$ kg & $0.457143$ sec & $21.9005$ sec & $0.458905$ sec  &  $1.81058$ sec & $8.99465$ sec \\
		$0.5$ nm & $6.0\times10^{-17}$ kg & $0.142857$ sec & $3.25995$ sec & $0.143125$ sec  &  $0.277719$ sec & $0.971052$ sec \\
		\midrule[0.3pt]\bottomrule[1pt]
	\end{tabular}
\end{table*} 
Since we are considering non-linear effects sufficiently small, in order to be consistent with the inequality in Eq. \eqref{eq:5} we need to require
\begin{equation}
2Gm^{3}\sigma_{0}\mathrm{Erfi}\left(\frac{M_s\sigma_{0}}{\sqrt{4+3M_{s}^{2}\sigma_{0}^{2}}}\right) <1, \label{eq:24}
\end{equation}
that can be seen as a quantifier of non-linearity, as the one in Eqs. \eqref{eq:5}-\eqref{eq:6}, and from which we can determine $\it again$ a threshold value for the mass:
\begin{equation}
\chi_{\scriptscriptstyle IDG}=\left[2G\sigma_{0}\mathrm{Erfi}\left(\frac{M_s\sigma_{0}}{\sqrt{4+3M_{s}^{2}\sigma_{0}^{2}}}\right)\right]^{-\frac{1}{3}}, \label{eq:25}
\end{equation}
that in the regime $M_s\sigma_{0}>2$ reduces to the one of Newton's gravity:
\begin{equation}
\chi_{\scriptscriptstyle N}=\left[2G\sigma_{0}\mathrm{Erfi}\left(\frac{1}{\sqrt{3}}\right)\right]^{-\frac{1}{3}}. \label{eq:26}
\end{equation}
If we choose the values $M_s=0.004$ eV and $\sigma_{0}=500$ nm for the scale of non-locality and the initial spread, respectively, we obtain $\chi_{\scriptscriptstyle IDG}\simeq 3.1\times 10^{-17}$ kg and $\chi_{\scriptscriptstyle N}\simeq 6.1\times 10^{-18}$ kg. These values are of the same order of the threshold masses $\mu_{\scriptscriptstyle IDG}$ and $\mu_{\scriptscriptstyle N}$ that we have found above by equating kinetic and gravitational energy contributions. Thus, Eq. \eqref{eq:24} is $\it consistent $ with the inequality in Eq. \eqref{eq:5}.

It is very clear by comparing Eqs. \eqref{eq:22}, \eqref{eq:23} and \eqref{eq:time-scale Newton} that 
\begin{equation}
\tau_{\scriptscriptstyle{\text{free}}}< \tau_{\scriptscriptstyle{IDG}}\leq \tau_{\scriptscriptstyle{N}}, \label{eq:29}
\end{equation}
namely the gravitational self-interaction causes a slow-down of the spreading of the wave-packet compared to the free-particle case. Moreover, since at short distances IDG interaction is weaker than the Newtonian one, a self-gravitating wake-packet in the IDG case will spread more quickly compared to Newton's theory.

We have solved numerically the integral in Eq. \eqref{eq:20} for both IDG and Newton's theory, and in Fig. \ref{fig2} we have plotted the radial probability density $\rho(\vec{x},t)=4\pi|\vec{x}|^{2}|\psi(\vec{x},t)|^2$ in the free, Newtonian and IDG cases at two fixed values of time as a function of the radial coordinate $|\vec{x}|$. We can immediately notice that the results in the two plots are in agreement with the analytical estimation made in Eq. \eqref{eq:29}.

Such gravitational inhibitions of the spreading, not only would offer a way to explore classical and/or quantum properties of the gravitational interaction, but would also provide a new framework to test short-distance gravity beyond GR, by investigating the real nature of the gravitational potential. 

\section{Discussion}
In the previous section we have found quantum spreading solutions for a self-gravitating wave-packet both in the case of IDG and Newtonian self-interactions. The results we have obtained, especially the analytical estimation in Eq. \eqref{eq:29} and the numerical solutions in Fig. \ref{fig2}, provide us a unique window of opportunity to test modified theories of gravity in a laboratory, in particular IDG-  as an example of singularity-free theory of gravity. 

Although the previous analysis holds for both cases of semi-classical and quantized gravity, in order to discuss the experimental testability of the model we need to distinguish between the two cases. As we have already mentioned above: 
\begin{enumerate}
	\item In the case of quantized gravitational interaction, Eq. \eqref{eq:1} describes the dynamics of a condensate, where the mutual gravitational interaction of all components would give an effective self-potential-contribution as result of the mean-field approximation; 
	\item In the semi-classical approach Eq. \eqref{eq:1} can describe the dynamics of a self-gravitating one particle system, as for instance elementary particles, or molecules whose center-of-mass' dynamics would be taken into account.
\end{enumerate}

The first case might be more interesting to explore: let us remind that the main motivations of IDG concern problems arising when trying to quantize the gravitational interaction, for example unitarity and renormalizability as already mentioned in Section II. Moreover, a semi-classical approach to IDG, would also imply a modification of quantum mechanics such that the fundamental equation governing the dynamics of a single-particle state would be non-linear. In this respect, case 1. seems less speculative than case 2..

However, in this section we wish to study the current and future experimental testability of our predictions about both classical and quantum aspects of gravity.

As far as case 1. is concern, there is a very promising experiment that is aimed to test quantum mechanics of weakly coupled Bose-Einstein condensates (BEC) in a freely falling system, where the spread of the quantum wave-packet would be tested in microgravity, see Ref. \cite{Muntinga:2013pta}. The used BEC was made of about $10^{4}$ atoms, but technology is progressing and in the near future it will be possible to consider BEC with $10^6$, or even more, atoms, allowing us to compare our predictions with the experimental data in such a way to constrain the scale of non-locality $M_s$, that so far has been only bounded in torsion-balance experiments \cite{Edholm:2016hbt}.

Regarding case 2., molecule-interferometry \cite{gerlich} seems to be one of the most favorable scenario to verify predictions of the semi-classical approach. Although it is not the configuration analyzed in this paper, it is also worth mentioning that another promising scenario aimed to test semi-classical gravity is given by optomechanics as explained in Refs. \cite{yang,andregro,grobardt-bateman}, where one considers many-body systems with a well-localized wave-function for the center of mass.  

All these kinds of experiments are very sensitive and, unfortunately, there are several sources of noise that need to be taken into account, as for example decoherence effects, see Ref. \cite{Bassi} for a review, and they represent a big challenge to overcome.

The choices of the initial spread $\sigma_{0}$ and the mass $m$ are very crucial in order to determine the time-scale for the spreading. First of all, to have appreciable gravity-induced effects we need values of the mass that are not much smaller than the threshold mass. Moreover, if we take $\sigma_{0}=500$ nm and $m\sim \chi_{\scriptscriptstyle N}\simeq 6.0\times 10^{-18}$ kg, the time-scales are of the order of $\tau_{\scriptscriptstyle{IDG}},\tau_{\scriptscriptstyle{N}}\sim \mathcal{O}(10^{5})$-$\mathcal{O}(10^{6})$ sec and $\tau_{\scriptscriptstyle{\text{free}}}\sim {\cal O}(10^{4})$ sec. In an interferometric experimental setup, for example, these values would not be suitable to test short-distance gravity, because the time-scales are too large compared to the coherence-times achieved by a modern matter-wave interferometer ($1 \text{-}3$ seconds) \cite{arndt-limits}. This means that we need smaller initial spreads so that the values of the time-scales will also decrease. For instance, if we choose $\sigma_{0}=1$ nm, $m=4.8\times 10^{-17}$ kg and $M_s=300$ eV, the time-scales turn out to be of the order of $\tau_{\scriptscriptstyle{\text{free}}}\sim \mathcal{O}(10^{-1})$ sec, $\tau_{\scriptscriptstyle{IDG}}\sim \mathcal{O}(1)$ sec and $\tau_{\scriptscriptstyle{N}}\sim \mathcal{O}(10)$ sec, which are more suitable values of time to test and compare modified theories of gravity.

In the table $\mathrm{I}$ we have shown some values of the three time-scales for fixed values of initial spreads and masses. It is very clear that by decreasing the initial spread, the time-scales also decrease. Those numerical values apply to both cases 1. and 2.; of course the setup can be different depending on the kind of experiments, and so the preparation of the initial state will also differ. In tests with BEC \cite{Muntinga:2013pta} the experimental configuration is given by an asymmetric Mach-Zehnder interferometer, while molecular-interferometry can be performed, for instance, with a Talbot-Lau interferometer \cite{gerlich}. In both cases the spread $\sigma_{\scriptscriptstyle 0}$ of the initial wave-packet is related in some way to the size of the slit separation $d$ in the interferometric setup. 

For example, as we have already mentioned in the footnote \ref{foot:1}, in a Talbot-Lau interferometry setup the initial spread $\sigma_{0}$ is of the same order of the slit separation $d$, that in turn is expressed in terms of the length $L$ of the interferometric device and of the the wave-length $\lambda$ of the particle: $L=d^{2}/\lambda$. The smallest slit separation that has been achieved so far is $d=500$ nm \cite{arndt2005}, which also implies $\sigma_{0}\sim 500$ nm. However, we have seen that in order to build a very suitable experimental scenario in which we can test modified theories of gravity we need at least a value $\sigma_{0}=1$ nm for the initial spread, which means that technology should decrease at least of two orders of magnitude the slit-separation $d$. Moreover, we also need masses of the order of $\mathcal{O}(10^{-18}\text{-}10^{-17})$ kg, and this is one of the biggest challenge to overcome: in fact, the most massive systems for which interference patterns have been observed are organic molecules with a mass of $10^{-22}$-$10^{-21}$ kg, see Ref. \cite{arndt2005}. In view of this last observation, it is worthwhile to mention that in Ref. \cite{pino} the authors present a new exciting proposal in which one might be able to perform quantum-interference with superconducting spheres with masses of the order of $10^{-14}$ kg, and in such a regime both gravitational and quantum effects should be not negligible.

In this manuscript we have extended the window of opportunity to test classical and quantum properties of modified gravity provided in Ref. \cite{l.buoninfante-solitonic}, where stationary properties of a quantum wave-packet were taken into account. Indeed, here we have made additional predictions that might allow us to test and constrain gravitational theories by studying non-stationary properties, in particular the quantum spreading of the wave-packet associated to a self-gravitating meso-scopic system in alternative theories beyond Einstein's GR. We have found a unique feature of IDG, as example of singularity-free theory of gravity, that predicts a faster spreading of the wave-packet compared to Newton's theory. A future observation of these predictions, even in a table-top experiment, might allow us a deeper and clearer understanding of short-distance gravity and let us learn more about its real nature, whether it is classical or quantum. \\

{\it Acknowledgements:} The authors thank the anonymous referees for constructive comments. The authors are also grateful to Steven Hoekstra and Claus L\"ammerzahl for discussions.

\end{document}